\documentclass{acm_proc_article-sp}
\usepackage{graphicx}

\conferenceinfo{}{Bloomberg Data for Good Exchange 2016, NY, USA}

\newcommand{\toolname}{VIOLA}
\newcommand{\MajVote}{MajVote}
\newcommand{\LabelReview}{LabelReview}
\newcommand{\vBasic}{Basic}
\newcommand{\vReview}{Review}
\newcommand{\vMultiBox}{MultiBox}
\newcommand{\vLabelDays}{LabelDays}
\newcommand{\vTracking}{Tracking}

\usepackage{color}
\usepackage{amssymb}
\usepackage{graphicx}
\usepackage{epstopdf}
\usepackage{subfig}
\usepackage{varwidth}
\usepackage{algorithm}
\usepackage{algpseudocode}
\usepackage{wrapfig}
\usepackage{array}
\usepackage{tabularx}

\usepackage{url}

\newcolumntype{C}[1]{>{\centering\arraybackslash}p{#1}}

\begin{document}

\title{Video Labeling for Automatic Video Surveillance in Security Domains}

\numberofauthors{8}
\author{
\alignauthor
Elizabeth Bondi, Debarun Kar, Venil Noronha, Donnabell Dmello, Milind Tambe\\
       \affaddr{University of Southern California}\\
       \affaddr{Los Angeles, CA}\\
       \email{bondi, dkar, vnoronha, ddmello, tambe@usc.edu}
\alignauthor
Fei Fang\\
       \affaddr{Carnegie Mellon University}\\
       \affaddr{Pittsburgh, PA}\\
       \email{feifang@cmu.edu}
\and
\alignauthor Arvind Iyer, Robert Hannaford\\
       \affaddr{AirShepherd}\\
       \affaddr{Berkeley Springs, WV}\\
       \email{arvind.iyer@lindberghfoundation.org, rob@coolideassolutions.com}
}

\maketitle

\begin{abstract}

Beyond traditional security methods, unmanned aerial vehicles (UAVs) have become an important surveillance tool used in security domains to collect the required annotated data. However, collecting annotated data from videos taken by UAVs efficiently, and using these data to build datasets that can be used for learning payoffs or adversary behaviors in game-theoretic approaches and security applications, is an under-explored research question. This paper presents \toolname, a novel labeling application that includes (i) a workload distribution framework to efficiently gather human labels from videos in a secured manner; (ii) a software interface with features designed for labeling videos taken by UAVs in the domain of wildlife security. We also present the evolution of \toolname~and analyze how the changes made in the development process relate to the efficiency of labeling, including when seemingly obvious improvements did not lead to increased efficiency. \toolname~enables collecting massive amounts of data with detailed information from challenging security videos such as those collected aboard UAVs for wildlife security. \toolname~will lead to the development of new approaches that integrate deep learning for real-time detection and response.

\end{abstract}

\section{Introduction} \label{sec:intro}

With the recent use of unmanned aerial vehicle (UAV) technology in security domains, videos taken by UAVs have become an emerging source of massive data \cite{hodgson2016precision}, especially in the domain of wildlife protection. 
For example, in security games, detecting wildlife from UAV videos can help estimate the animal distribution density, which decides the payoff structure of a security game. Detecting poachers and their movement patterns could also lead to successful learning of attackers' behavioral models, which is an important topic in security games \cite{Nguyen13a:SUQR,Kar15a}. 
In addition, data collected from UAVs can enable the development of a new generation of game-theoretic tools for security. Particularly, the data can be used to train or fine-tune a deep neural network to automatically detect attackers from the video taken by the UAVs in real-time.

Unfortunately, collecting labeled data from videos taken by UAVs can be a labor-intensive, time-consuming task. To our knowledge, there is no existing application that focuses on assisting the labeling work for videos taken by UAVs in security domains. Existing applications for labeling images \cite{ImageNet,Everingham10} cannot be directly applied to labeling videos, as treating each frame as a separate image can lead to inefficiency, since it does not exploit the correlation between frames. Video labeling applications such as VATIC \cite{VATIC} attempt to choose key frames for labeling, or track objects through the video. However, in UAV videos with camera motion, possibly collected using a different wavelength, these methods may not apply and may lead to inaccurate results or extra work for labelers, since the position of the objects in the video may change abruptly and the lack of color bands makes the tracking much more difficult. Furthermore, these applications are often paired with AMT to get labeled video datasets from online workers. However, in a security domain with sensitive data, meaning data that would provide attackers with some knowledge of defenders' strategies should it be shared, it may be undesirable to use AMT. This would then require finding labelers and organizing labeling assignments. 

In this paper, we focus on better collection of labeled data from UAVs to provide input for game-theoretic approaches for security, and in particular to security game applications for wildlife conservation such as PAWS \cite{fang2016deploying}. There has been work on labeling tools in domains such as computer vision and cyber security \cite{ImageNet,catania2012autonomous}, but there exists no work on labeling tools for game-theoretic approaches. 

In particular, we will focus on labeling videos taken by long wave thermal infrared (hereafter referred to as thermal infrared) cameras installed on UAVs, in the domain of wildlife security. We present \toolname~(VIdeO Labeling Application), a novel application that assists labeling objects of interest such as wildlife and poachers. \toolname~includes a workload distribution framework to efficiently gather human labels from videos in a secured manner. We distribute the work of labeling the videos and reviewing the labels amongst a small group of labelers to ensure efficiency and data security. \toolname~also provides an easy-to-use interface, with a set of features designed for UAV videos in the wildlife security domain, such as allowing for moving multiple bounding boxes simultaneously and tracking bright regions automatically. We will also discuss the various stages of development to create \toolname, and we will analyze the impact of different labeling procedures and versions of the labeling application on efficiency, with a particular emphasis on the surprising results that showed some changes did not improve efficiency.
\section{Related Work}
Game-theoretic approaches have been widely used in infrastructure and green security domains \cite{tambe11}. In green security domains such as protecting wildlife from poaching, multiple research efforts in artificial intelligence and conservation biology have attempted to estimate wildlife distribution and poacher activities \cite{fang2016deploying}; such efforts often rely on months or years of recorded data \cite{nguyen2016capture,Kar:2017:INTERCEPT}. With the recent advances in unmanned aerial vehicle (UAV) technology, there is an opportunity to provide detailed data about wildlife and poachers for game-theoretic approaches. Since a poacher is rewarded for successfully poaching wildlife, the wildlife distribution determines the payoff structure of the game. Poachers' behavioral models can be inferred from poaching activities and be used to design better patrol strategies with game-theoretic reasoning. In addition, game-theoretic patrolling with alarm systems \cite{alpcan2003game,Basilico2016} has been studied. UAVs can provide input for such systems in real-time using computer vision, particularly by detecting attackers or suspicious human beings in the UAV videos. 

Detecting attackers in the UAV videos is related to object detection. Recently, great progress has been achieved in computer vision by deep learning in object detection and recognition \cite{FasterRCNN,YOLO}. However, state-of-the-art detectors cannot be directly applied to our aerial videos because most methods focus on detection in high resolution, visible spectrum images. An alternative approach to this detection is to track moving objects throughout videos. Tracking of both single and multiple objects in videos has been studied extensively \cite{yilmaz2006object}. These methods also rely on high resolution visible spectrum videos. Single object trackers use discriminant features from high resolution videos to establish correspondences \cite{kristan2015visual}. Much of multi-object tracking research is directed towards pedestrians \cite{bae-mot1,zhang-mot2,mot16}, and primarily focuses on visible spectrum videos with high resolution, or videos taken from a fixed camera (except \cite{mot16}).

Simpler and more general tracking algorithms exist that do not necessarily have these dependencies, such as the Lucas-Kanade tracker for optical flow \cite{lucas-kanade}, popular in the OpenCV package, and general correlation-based tracking \cite{correlation}. Small moving objects can also be detected by a background subtraction method after applying video stabilization \cite{pai2007moving}. Because these methods are more general, they are still applicable to our domain and were explicitly tested, but still did not perform well in many cases. For example, since the video stabilization and background subtraction method assumes a planar surface, in the case of more complex terrain, there were many noisy detections. Instead of using tracking for detection, we therefore decided to focus on deep learning. 

In order to use deep learning-based detection methods with aerial, thermal infrared data, hand-labeled training data are required to fine-tune the networks or even train them from scratch. In addition to video labeling applications such as VATIC \cite{VATIC}, there has been work on semi-automatic labeling \cite{Yan03} and label propagation \cite{badrinarayanan10} which combines the effort of human labelers and algorithms to speed up the labeling process for videos. This work often focuses on how to select the frames for human labelers to label and how to propagate the labels for the remaining frames. This is difficult for our domain because of the motion of UAVs, and because it is often hard for humans to tell which objects are of interest without seeing the object's motion. As a result, we sought to develop our own labeling application, \toolname. The first key component of the application is a workload distribution framework. A common framework for image and video labeling is a majority voting framework \cite{Nguyen13,Park2012,Nguyen-Dinh13,Sheng08}. \toolname~uses a framework based upon \cite{Everingham10} to efficiently gather labels from a small group of labelers. We examine the framework further in Sec. \ref{development} and Sec. \ref{analysis}.
\section{Domain} \label{data}
There has recently been increased use of UAVs for security surveillance. UAVs are able to cover more ground than a stationary camera and can provide the defenders more advanced notice of a potential threat. To detect suspicious human activities at night, the UAVs can be equipped with thermal infrared cameras. This is the type of UAV video we deal with in our domain, since poaching often occurs at night. We will specifically be able to use these types of data to detect poachers and provide advanced notice to park rangers, and use these detections to provide input for patrol generation tools such as PAWS.

\begin{figure}[t]
	\includegraphics[width=0.3\textwidth, height=4cm]{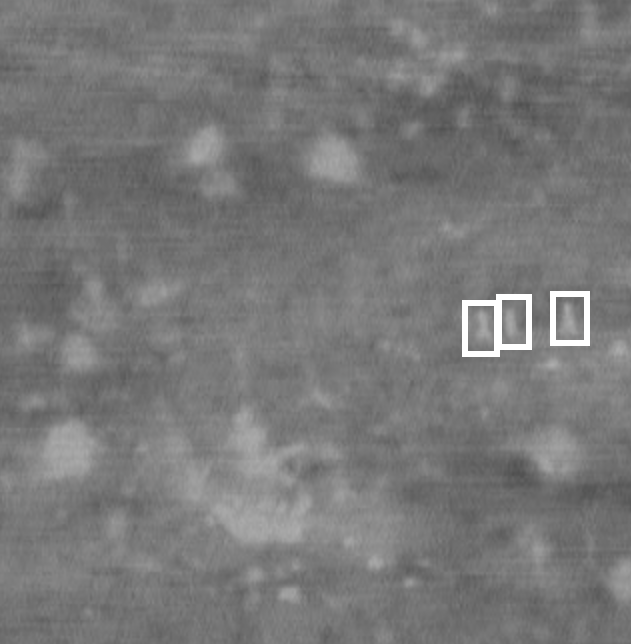}
	\centering
	\caption{An example of a thermal infrared frame, where the three humans outlined by the white boxes look very similar to the surrounding vegetation.}
	\label{fig:similarities_example}
\end{figure}

In order to accomplish this, we need labeled data from the thermal infrared, UAV videos in the form of rectangular ``bounding boxes'' for objects of interest (animals and poachers) in each frame, with a color corresponding to their classification. However, the movement of UAVs and the thermal infrared images make it extremely difficult to label videos in this domain. First, thermal infrared cameras are low-resolution, and typically show warmer objects as brighter pixels in the image, although the polarity could be reversed occasionally. Different phenomena could also cause brighter pixels without a warm object. For example, the ground warms during the day, and then emits heat at night, which can be reflected under a tree canopy and lead to an amplified signal that might look like a human or animal. Furthermore, vegetation often looks bright and similar to objects of interest, as in Fig. \ref{fig:similarities_example}, where there are three humans labeled with bounding boxes, amongst many other bright objects. Second, since the data are captured aboard a moving UAV, these data often vary drastically. For example, the resolution, and therefore size of targets, is very different throughout the dataset because the UAV flies at varying altitudes.

\begin{figure}[t]
	\includegraphics[width=0.23\textwidth]{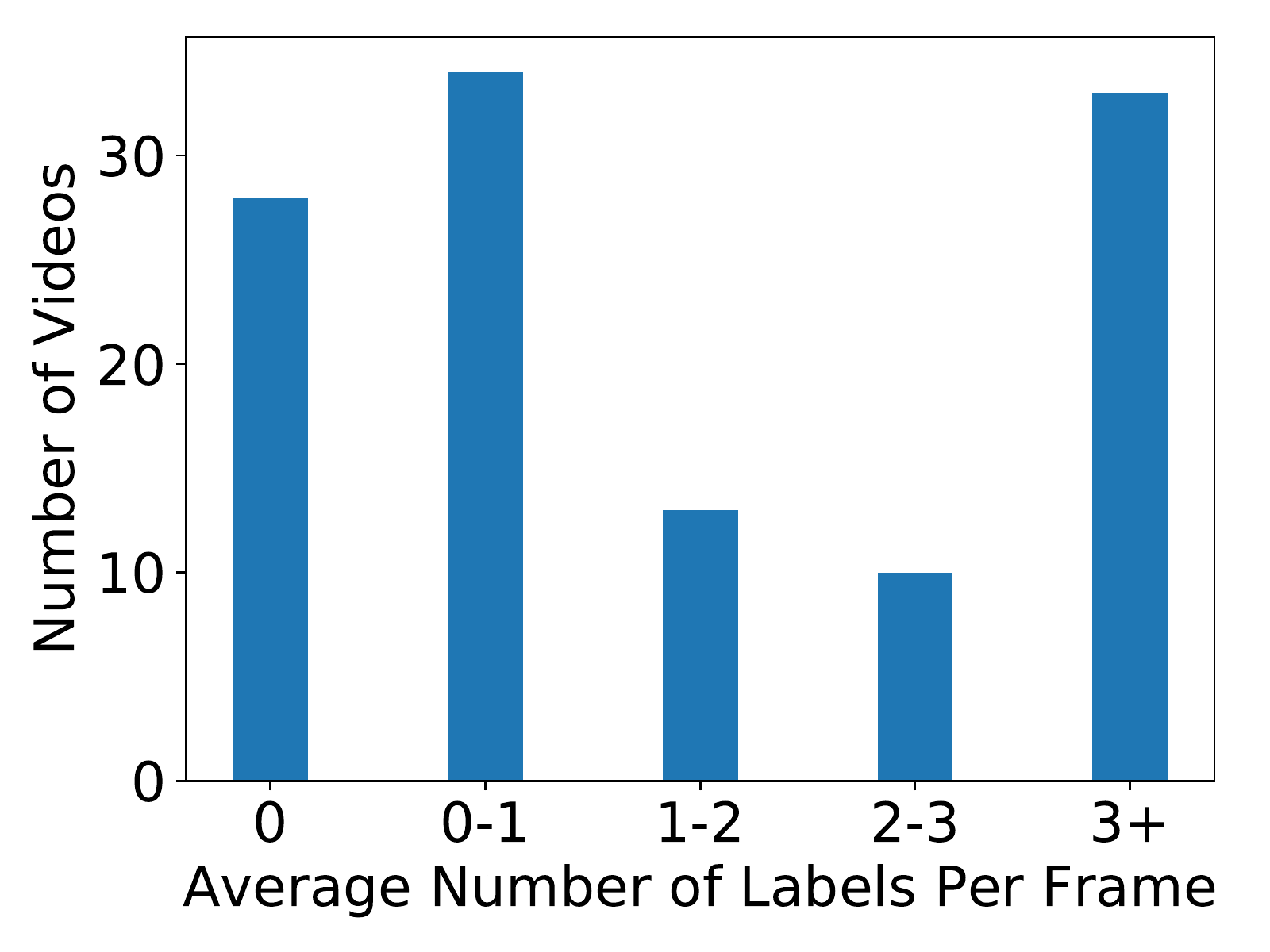}
	\includegraphics[width=0.23\textwidth]{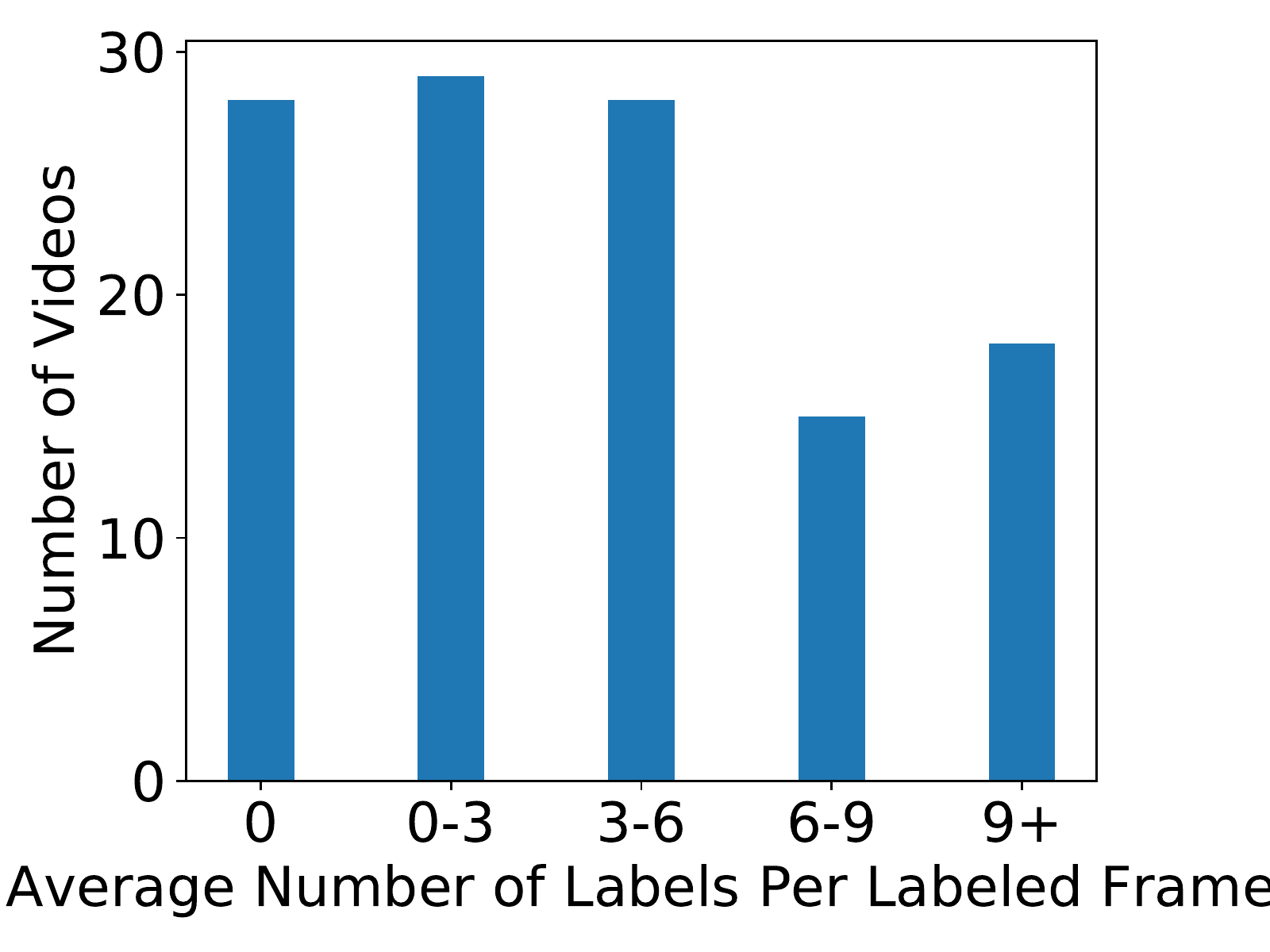}
	\centering
	\caption{A histogram with the number of videos for average objects of interest per frame (left), and the average objects of interest per labeled frame (right).}
	\label{fig:histogram}
	\vspace{-10pt}
\end{figure}

In addition to difficult, variable video data to begin with, some videos may have many objects of interest in them, whereas some videos may not have any objects of interest at all. It sometimes takes a long time to determine if there are any objects of interest, and it also often takes a long time to label when there are many objects of interest. To illustrate the variation in the number of objects of interest, we analyze the historical videos we get from our collaborators. Fig. \ref{fig:histogram} shows a histogram of the average number of labels per frame, meaning that all frames in the video were counted, regardless of whether or not they were labeled, and a histogram of the average number of labels per labeled frame, meaning frames that had at least one label were counted.

Although we focus on UAV videos in wildlife security domains, similar challenges in UAV videos in other security domains can be expected. Therefore, the application \toolname~we introduce in this paper can potentially be applied to other security domains to provide input for game-theoretic approaches.
\section{\toolname} \label{basic-framework}

The main contribution of this paper is \toolname, an application we developed for labeling UAV videos in wildlife security domains. 
\toolname~includes an easy-to-use interface for labelers and a basic framework to enable efficient usage of the application. In this section, we first discuss the user interface and then the framework for work distribution and training process for labelers.

\subsection{User Interface of \toolname} \label{app}

The user interface of \toolname~was written in Java and Javascript, and hosted on a server through a cloud computing service so it could be accessed using a URL from anywhere with an internet connection. 

\begin{figure}[t]
	\subfloat[][]{%
		\label{fig:menus1}%
		\includegraphics[width=0.23\textwidth, height=3cm]{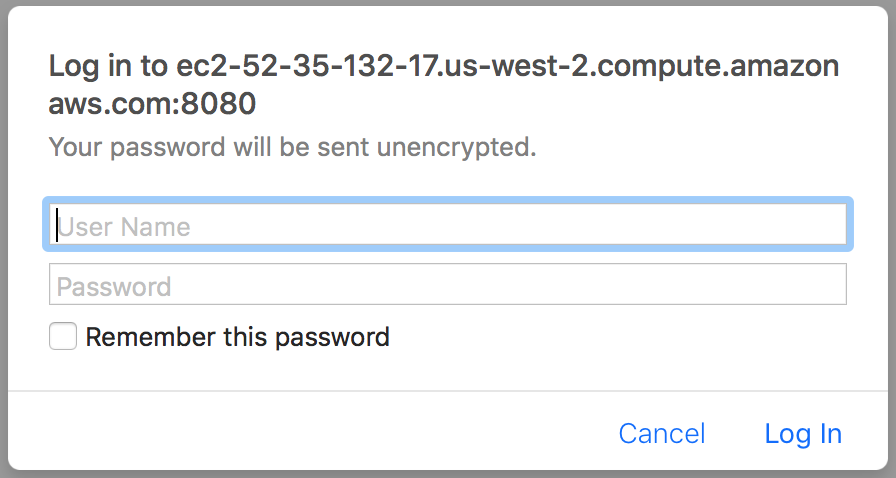}}%
	\subfloat[][]{%
		\label{fig:menus2}%
		\includegraphics[width=0.23\textwidth, height=3cm]{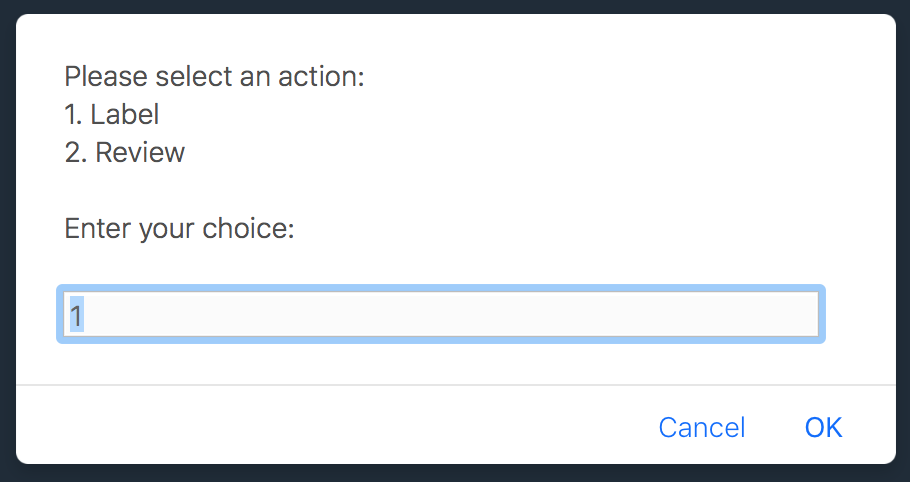}}\\
	\subfloat[][]{%
		\label{fig:menus3}%
		\includegraphics[width=0.23\textwidth, height=3cm]{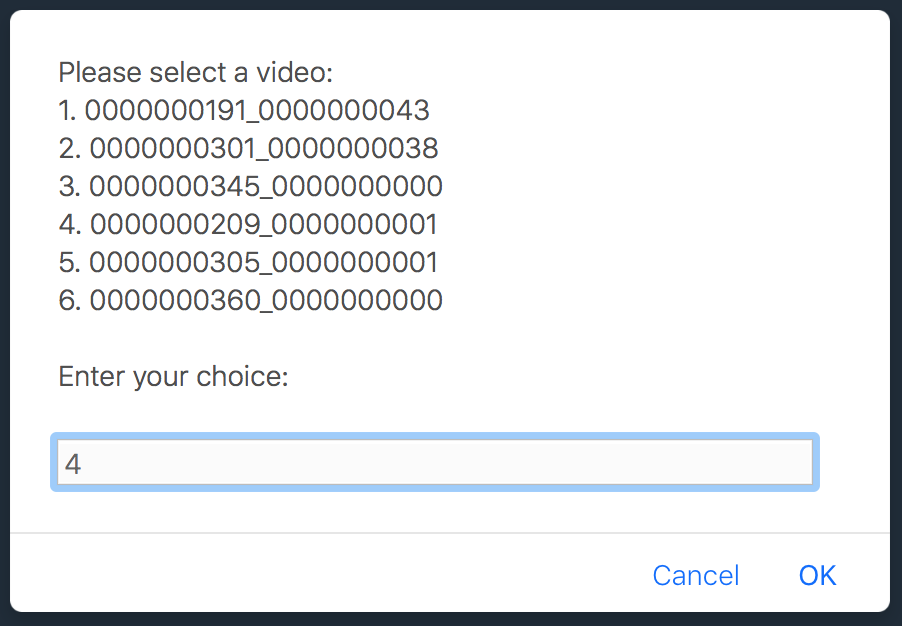}}%
	\subfloat[][]{%
		\label{fig:menus4}%
		\includegraphics[width=0.23\textwidth, height=3cm]{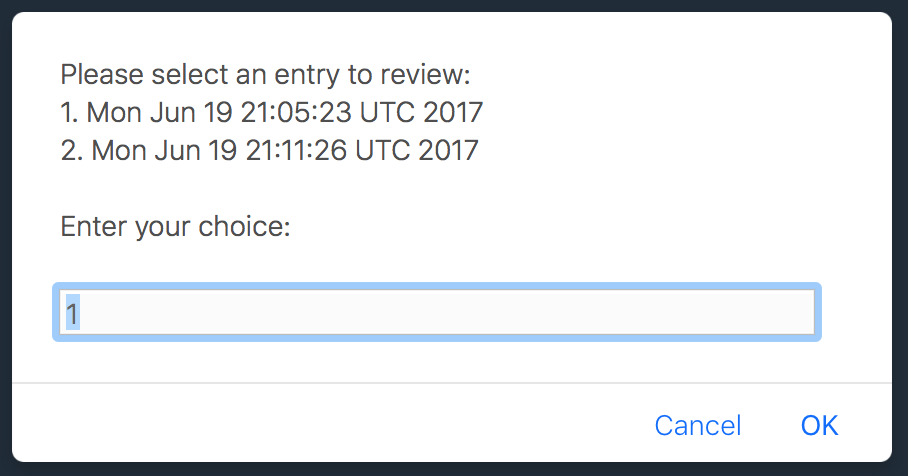}}%
	\centering
	\caption{The menus to begin labeling.}
	\label{fig:menus}
\end{figure}

\begin{figure}[t]
	\includegraphics[width=0.35\textwidth, height=4.5cm]{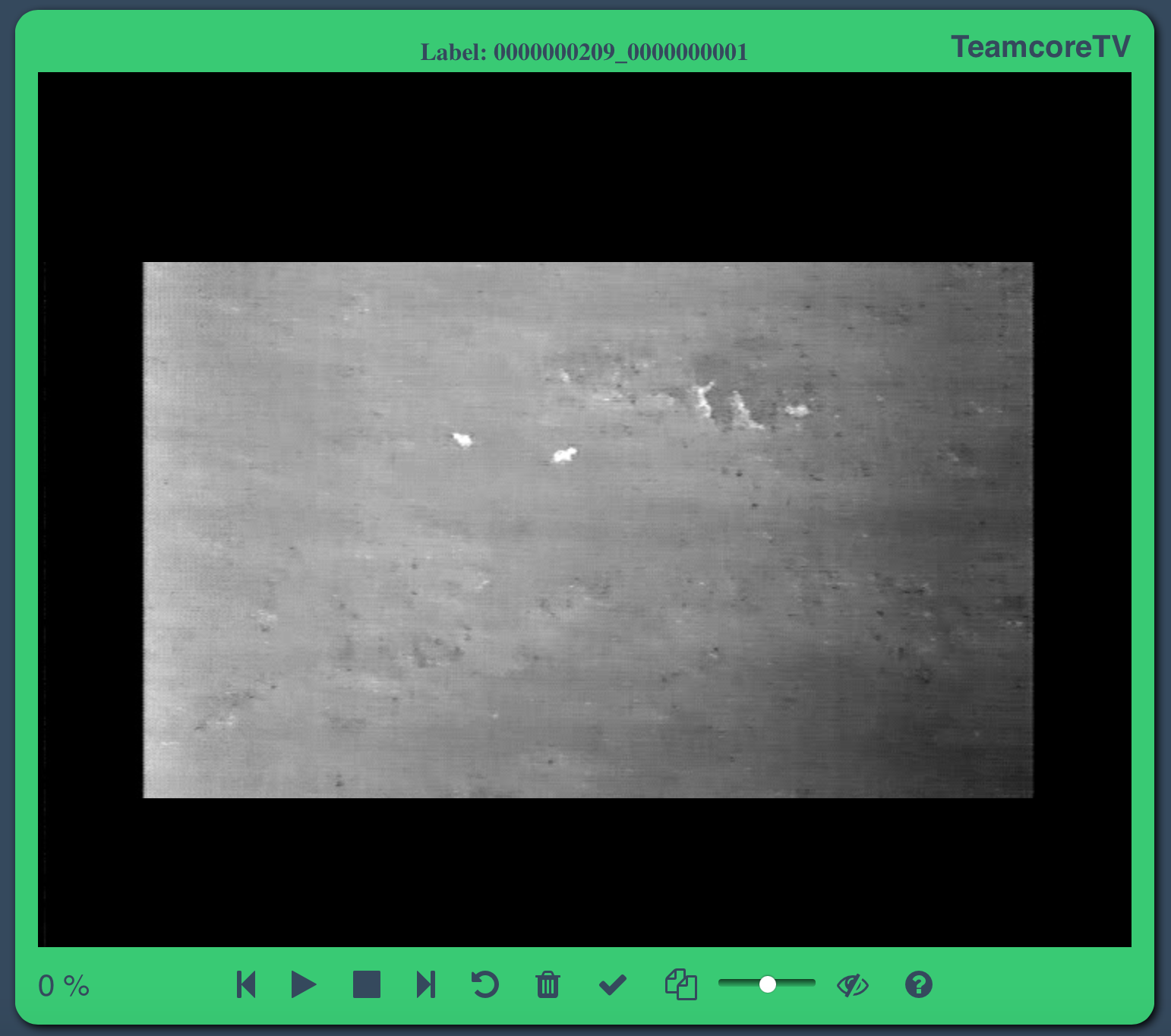}
	\centering
	\caption{An example of a frame (left) and labeled frame (right) in a video. This is the next screen displayed after all of the menus and allows the labeler to navigate through the video and manipulate or draw bounding boxes throughout.}
	\label{fig:example}
	\vspace{-10pt}
\end{figure}

Before labeling, labelers were asked to login to ensure data security (Fig. \ref{fig:menus1}). The first menu that appears after login (Fig. \ref{fig:menus2}) asks the labeler which mode they would like, whether they would like to label a new video or review a previous submission. Then, after choosing ``Label", the second menu (Fig. \ref{fig:menus3}) asks them to choose a video to label. Fig. \ref{fig:example} is an example of the next screen used for labeling, also with sample bounding boxes that might be drawn at this stage. Along the top of the screen is an indication of the mode and the current video name, and along the bottom of the screen is a toolbar. First, in the bottom left corner, is a percentage indicating progress through the video. Then, there are four buttons used to navigate through the video. The two arrows move backwards or forwards, the play button advances frames at a rate of one frame per second, and the square stop button returns to the first frame of the video. The next button is the undo button, which removes the bounding boxes just drawn in the current frame, just in case they were too tiny to easily delete. Also to help with the nuisance of creating tiny boxes by accident while drawing a new bounding box or while moving existing bounding boxes, there is a filter on bounding box size. The trash can button deletes the labeler's progress and takes them back to the first menu after login (Fig. \ref{fig:menus2}). Otherwise, work is automatically saved after each change and re-loaded each time the browser is closed and re-opened. The application asks for confirmation before deleting the labeler's progress and undoing bounding boxes to prevent accidental loss of work. The check-mark button is used to submit the labeler's work, and is only pressed when the whole video is finished. Again, there is a confirmation screen to avoid accidentally submitting half of a video. The copy button and the slider will be described further in Sec. \ref{development}. The eye button allows the labeler to toggle the display of the bounding boxes on the frame, which is often helpful during review to check that the labels are correct. Finally, the question mark button provides a help menu with a similar summary of the controls of the application. Notice the bounding boxes surrounding the animals in this video are colored red. Humans would be colored blue. This is also included in the help menu. 

To draw bounding boxes, the labeler can simply click and drag a box around the object of interest, then click the box until the color reflects the class. Deleting a bounding box is done by pressing SHIFT and click, and selecting multiple bounding boxes is done by pressing CTRL and click, which allows the labeler to move multiple bounding boxes at once. Finally, while advancing frames, bounding boxes drawn in the current frame are moved to the next frame. It only happens the first time a frame is viewed since it could otherwise add redundant bounding boxes or replace the bounding boxes originally added by the labeler.

If ``Review" is chosen in the first menu after login, the second menu also asks the labeler to choose a video to review, and then a third menu (Fig. \ref{fig:menus4}) asks them to choose a labeling submission to review. It finally displays the video with the labels from that particular submission, and they may begin reviewing the submission. The two differences between the labeling and review modes in the application are (i) that the review mode displays an existing set of labels and (ii) that labels are not moved to the next frame in review mode. 

\subsection{Use of \toolname} \label{use}

Our goal in labeling the challenging videos in the wildlife security domain is first to keep the data secure, and second, to collect more usable labels to provide input for game-theoretic tools for security. In addition, we aim for exhaustive labels with high accuracy and consistency. To achieve these goals, we securely distribute the work among a small group of labelers, assign labelers to either provide or review others' labels, and provide guidelines and training for the labelers.

\textbf{Distribution of Work} 
To keep the data (historical videos from our collaborators) secure, instead of using AMT, we recruit a small group of labelers, in this work 13. Labelers are given a username and password to access the labeling interface, and the images on the labeling interface cannot be downloaded. 

In order to achieve label accuracy, we use a framework of label and review. The idea is simply that one person labels a video, and another person checks, or reviews, the labels of the first person. By checking the work of the labeler, the reviewer must agree or disagree with the original set of labels instead of creating their own. Upon disagreement, the reviewer can change the original labels. This was primarily chosen because it is a clean approach, leading to one set of final labels per video. This choice is described further in Sec. \ref{development}. Assignments are organized using shareable spreadsheets.

\textbf{Guidelines and Training for Labelers}
In order to achieve accuracy and consistency of labels, we provide guidelines and training for the labelers. During the training, we show the labelers several examples of the videos and point out the objects of interest. We provide them with general guidelines on how to start labeling a video, as below (review is similar).

\textit{In general, the process for labeling should be:}
\begin{itemize}
	\item{Watch the video once all the way through and try to decide what you see.}
	\item{Once you have an idea of what is happening in the video by going through it, return to the beginning of the video and start labeling.}
	\item{Make and move bounding boxes.}
	\item{Send screenshots if you need help.}
\end{itemize}

We provide special instructions for the videos in our domain. For example, animals tend to be in herds, obviously shaped like animals, and/or significantly brighter than the rest of the scene, and humans tend to be moving. 
\section{Development} \label{development}

Thanks in large part to feedback provided by the labelers, we were able to make improvements throughout the development of the application to the current version discussed in Sec. \ref{app}. In the initial version of the application, we had five people label a single video, and then automatically checked for a majority consensus among these five sets of labels. We used the Intersection over Union (IoU) metric to check for overlap with a threshold of 0.5 \cite{Everingham10}. If at least three out of five sets of labels overlapped, it was deemed to be consensus, and we took the bounding box coordinates of the first labeler. Our main motivation for having five opinions per video was to compensate for the difficulty of labeling thermal infrared data, though we also took into account the work of \cite{Nguyen13} and \cite{Park2012}. 
The interface of the initial version allowed the user to draw and manipulate bounding boxes, navigate through the video, save work automatically, and submit the completed video. Boxes were copied to the next frame and could be moved individually. To get where we are today, the changes were as listed in Table \ref{tab:changes}.

\begin{table}[t]
\caption{Changes made throughout development.}
\label{tab:changes}
\centering
\begin{tabularx}{.47\textwidth}{|l|X|X|l|}
 \hline
 Version & Change & Date & Brief Description \\ 
 \hline
 1 & - & - & Draws and edits  \\
  & & & boxes, navigates  \\
  & & & video, copies \\
  & & & boxes to next \\
  & & & frame\\
 \hline
 2 & Multiple & 3/23/17 & Moves multiple \\
  & Box & & boxes at once to \\ 
  & Selection & & increase labeling \\ 
  & & & speed \\
  \hline
 3 & Five & 3/24/17 & Requires only two \\
  & Majority & & people per video \\
  & to & & instead of five to \\ 
  & Review & & improve overall \\
  & & & efficiency \\
  \hline
 4 & Labeling & 4/12/17 & Has labelers \\
  & Days & & assemble to \\
  & & & discuss difficult \\
  & & & videos \\ 
  \hline
 5 & Tracking & 6/17/17 & Copies and \\
  & & & automatically  \\
  & & & moves boxes to \\
  & & & next frame \\ 
 \hline
\end{tabularx}
\end{table}

The most significant change made during the development process was the transition from five labelers labeling the same video and using majority voting to get the final labels (referred to as ``\MajVote'') to having one labeler label the video followed by a reviewer reviewing the labels (referred to as ``\LabelReview''). We realized that having five people label a single video was very time consuming, and the quality of the labels was still not perfect because of the ambiguity of labeling thermal infrared data, which led to little consensus. Furthermore, when there was consensus, there were three to five different sets of coordinates to consider. Switching to \LabelReview~eliminated this problem, providing a cleaner and also time-saving solution.
Another change, ``Labeling Days", consisted of meeting together in one place for several hours per week so labelers were able to discuss ambiguities with us or their peers during labeling. 
Finally, the tracking algorithm (Alg. \ref{tracking}) was added to automatically track the bounding boxes when the labeler moves to a new frame to improve labeling efficiency, as the labelers would be able to label a single frame, then check that the labels were correct.

\begin{algorithm}[t]
\centering
\footnotesize
\caption{Basic Tracking Algorithm}
\label{tracking}

\begin{varwidth}{\dimexpr\linewidth-2\fboxsep-2\fboxrule\relax}
\begin{algorithmic}[1]
\State $buffer \gets userInput$
\ForAll {$boxesPreviousFrame$}
	\If {$boxSize > sizeThreshold$}
		\State $newCoords \gets coords$
	\Else	
		\State $searchArea \gets newFrame[coords + buffer]$
		\State $threshIm \gets \Call{Threshold}{$searchArea, thresh$}$
		\State $components \gets \Call{CCA}{$threshIm$}$
		\If {$numberComponents > 0$}
			\State $newCoords \gets \Call{GetLargest}{components}$		
		\Else
			\State $newCoords \gets coords$
		\EndIf
	\EndIf
	\State \Call{CopyAndMoveBox}{$newFrame, newCoords$}
	\EndFor
\end{algorithmic}
\end{varwidth}%
\end{algorithm}

\begin{figure*}[t]
	\centering
	\subfloat[][]{%
		\label{fig:process5}%
		\includegraphics[width=0.23\textwidth]{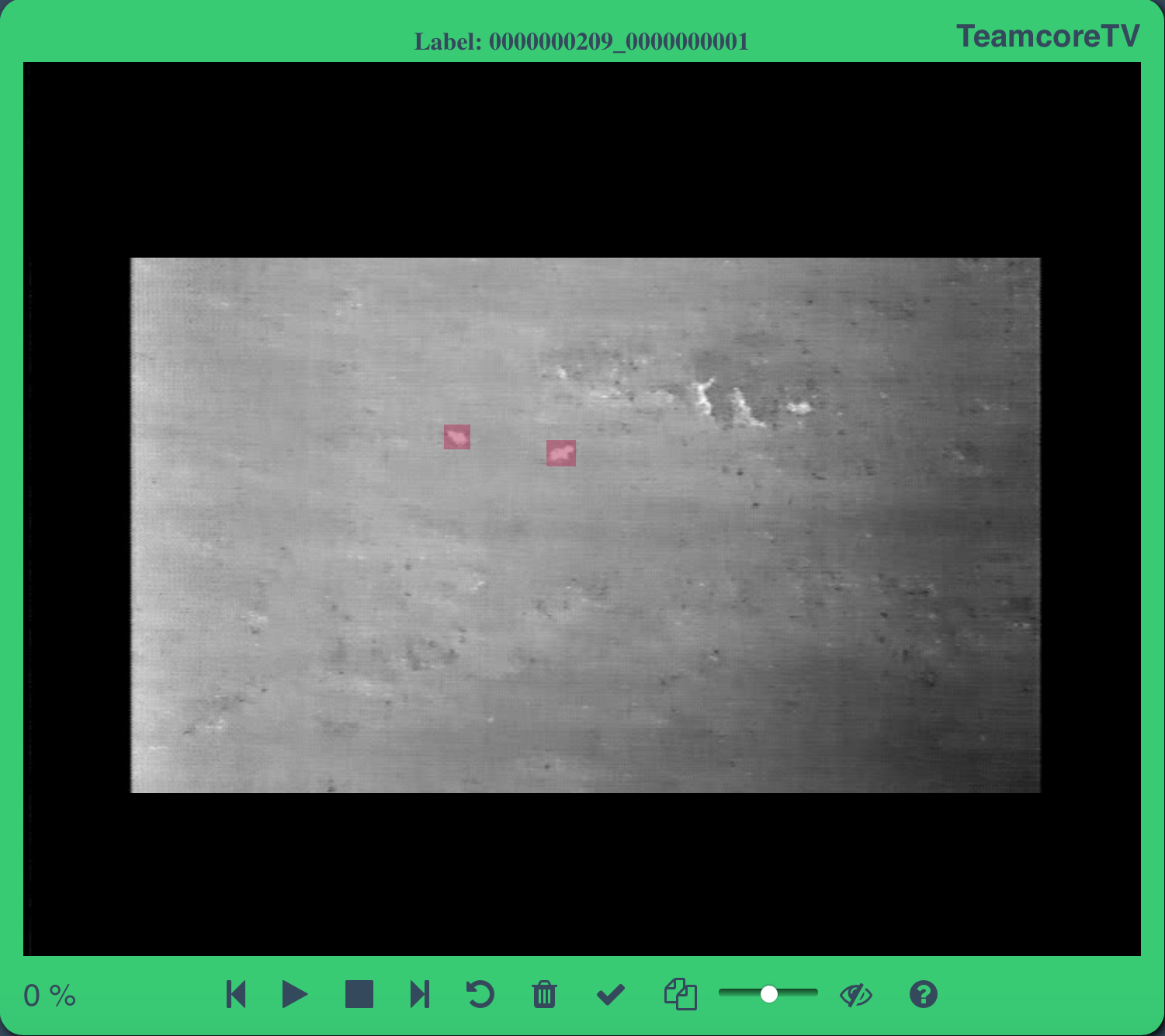}}%
	\subfloat[][]{%
		\label{fig:process6}%
		\includegraphics[width=0.23\textwidth]{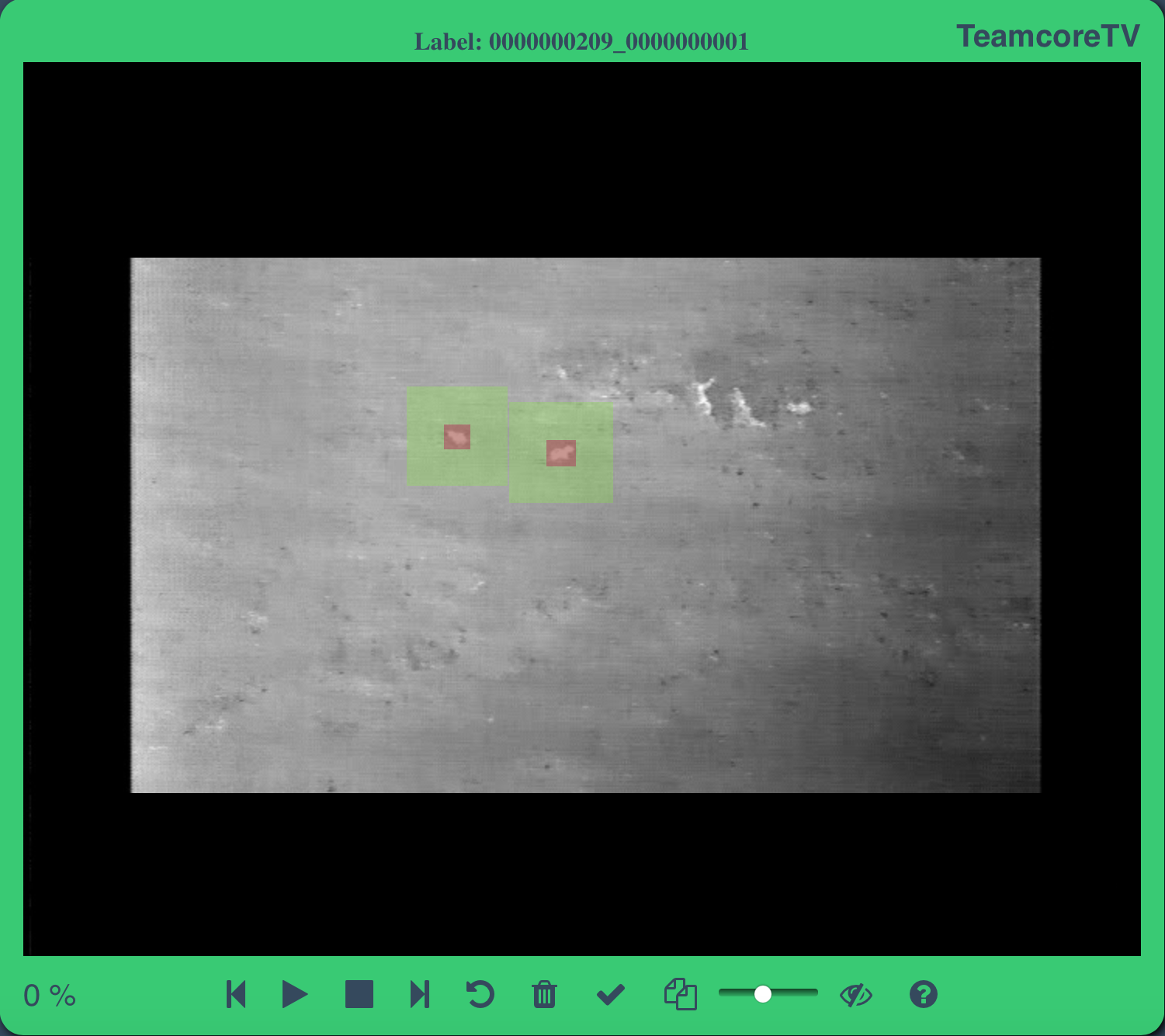}}%
	\subfloat[][]{%
		\label{fig:process7}%
		\includegraphics[width=0.23\textwidth]{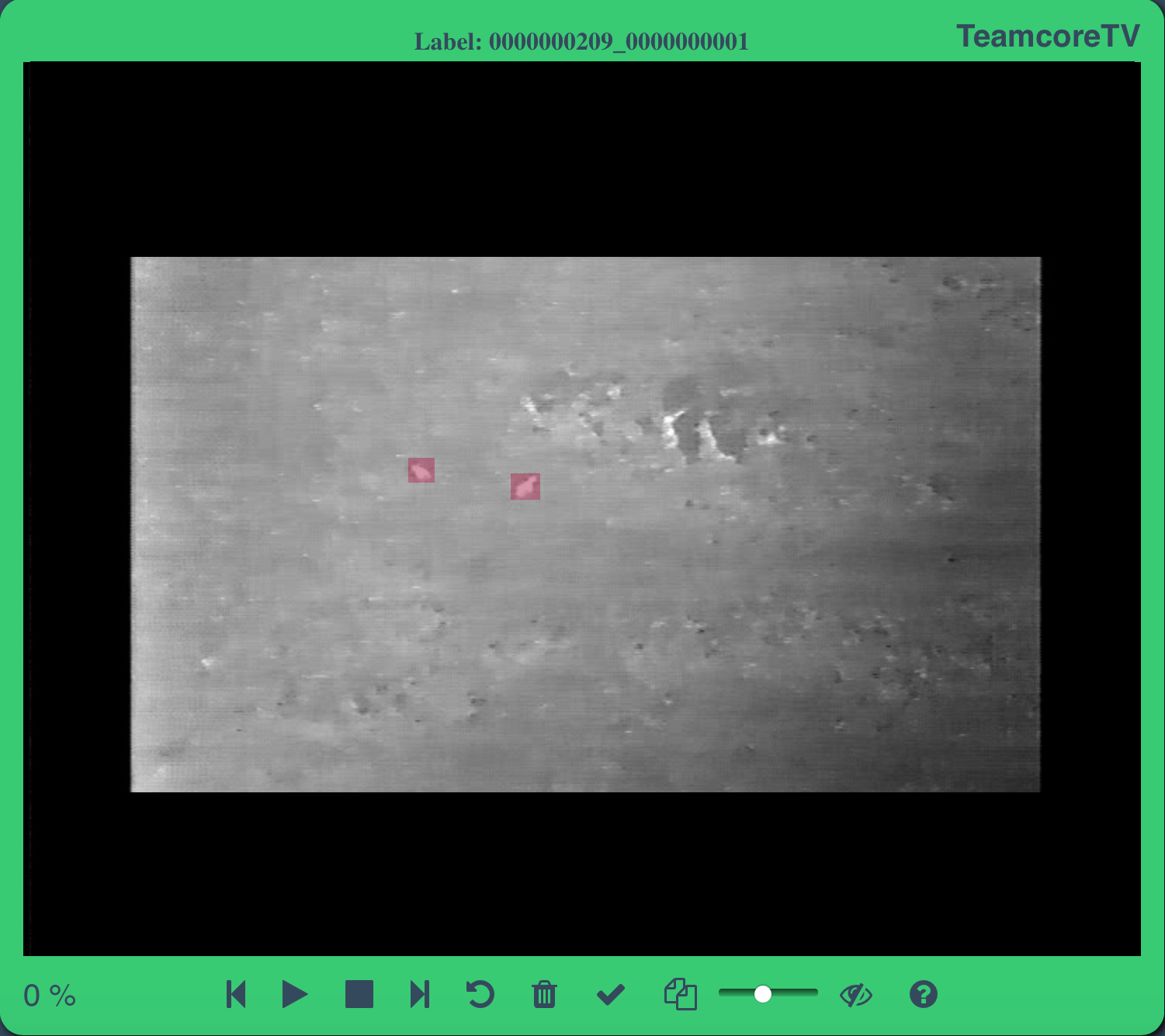}}%
	\subfloat[][]{%
		\label{fig:process8}%
		\includegraphics[width=0.23\textwidth]{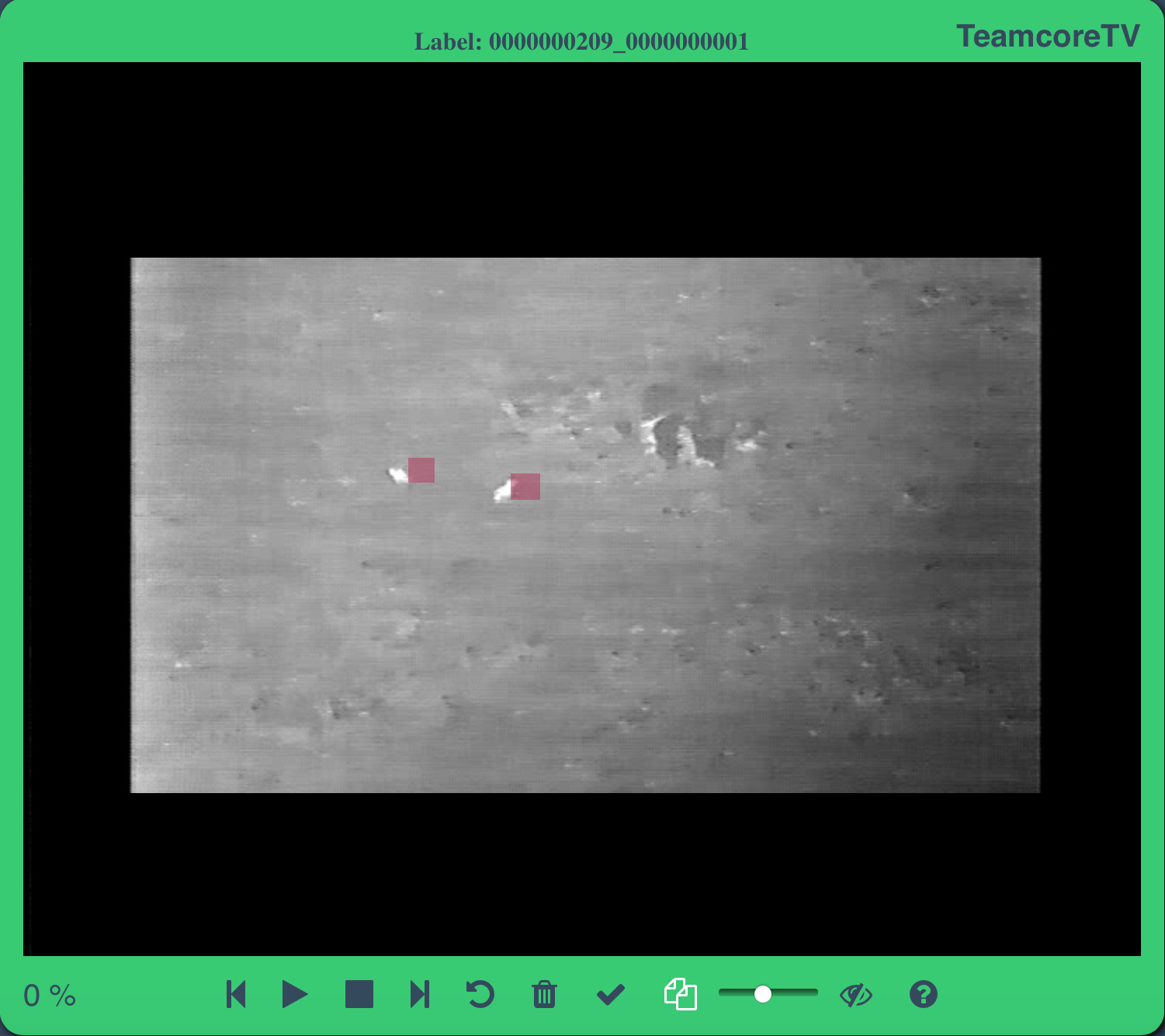}}%
	\centering
	\caption{A sample labeling process.}
	\label{fig:process}
\end{figure*}

An example of the tracking process in use is shown in Fig. \ref{fig:process}. First, the labeler drew two bounding boxes around the animals (Fig. \ref{fig:process5}), then adjusted the search size for the tracking algorithm using the slider in the toolbar (Fig. \ref{fig:process6}). The tracking algorithm was applied to produce the new bounding box location (Fig. \ref{fig:process7}). In contrast, the copy feature, activated when the copy button was selected on the toolbar, only copied the boxes to the same location (Fig. \ref{fig:process8}). In this case, since there was movement, and the animals were large and far from one another, the tracking algorithm correctly identified the animals in consecutive frames. If several bright objects were in the search region, it could track incorrectly and copying could be better. One direction of future work is to improve the tracking algorithm by setting thresholds automatically and accounting for close objects.

\section{Analysis}  \label{analysis}

In this section, we analyze how the changes we made during the development of \toolname~affect the labeling efficiency by examining two questions: (i) how the changes affect the overall efficiency of the data collection process, which is measured by the total person time needed to get a final label -- a label confirmed by the five majority voting or review that can be used for game-theoretic analysis or deep learning algorithms; (ii) how the changes affect the individual efficiency, i.e., the person time needed for an individual labeler or reviewer to provide or to check a label. In addition, we examine whether other desired properties of the data collection process, such as exhaustiveness, have been achieved.

To analyze efficiency, we first went through the person time data collected during the development of \toolname. Any changes made in \toolname~were deployed immediately to make faster progress in labeling the videos. These person time data came from different videos and labelers. They inherently took different amounts of time to label regardless of the application, since the videos varied in their content. To mitigate the intrinsic heterogeneity for analysis, we divide the videos into four groups, $(0, 1)$, $[1, 2)$, $[2, 3)$, and $[3, +\infty)$, based on the average number of labels per frame, since it is an important indicator of the difficulty of labeling a video. There were other factors affecting the difficulty of labeling videos, so videos in the same group may still have had high variation. Because of this, we remove the top and bottom 5\% of time per label entries.

Also due to these concerns, we collected additional person time data in a more controlled environment. We gave six unique videos that contained animals but no poachers to the labelers to label. The labelers had not seen these videos previously. We distributed the work among the labelers so as to get one set of final labels for each video under each of the versions of \toolname~(as shown in Table \ref{tab:changes}). We asked the labelers to label for no more than 15 minutes on each video. To accommodate the labelers' schedules and coordinate their schedules to set up meetings, which were necessary for  \vLabelDays~and \vTracking, we gave the labelers 2 to 4 days to label the videos under each version. As such, it was difficult to get multiple sets of labels for each video or get labels for more videos. Labelers may not have checked all the frames in the video within 15 minutes, so we use the minimum checked frame among labelers for each video under each version, and analyze efficiency using person time data up until that frame only. 
Also, note that since some labelers were asked to label the same video multiple times under different versions, the labelers likely got faster as time went on. To mitigate these effects, we randomly ordered the five versions of \toolname~for them to label. The order is shown in Table \ref{tab:order}. Framework was abbreviated as FW, \vMultiBox~as MB, \vLabelDays~as LD,  \vTracking~as Track, and \LabelReview~as LR in Table \ref{tab:order}.
\begin{table}[t]
\caption{Versions tested in additional tests.}
\label{tab:order}
\centering
\begin{tabularx}{.47\textwidth}{|l|X|X|X|X|l|}
 \hline
 Version & 1 & 2 & 3 & 4 & 5\\
\hline
 Name & \vBasic & MB & \vReview & LD & Track \\
 \hline
 FW & \MajVote & \MajVote & LR & LR & LR \\
 \hline
 Order & 4th & 3rd & 1st & 2nd & 5th \\
 \hline
 \end{tabularx}
\end{table}

We will proceed in this section by first focusing on the impact of the key change in the labeling framework from \MajVote~to \LabelReview~on the overall efficiency. We will then check each version of \toolname~to understand the impact of other changes. Because of the surprising results, we will also examine videos in which these features did not help.

\subsection{From \MajVote~to \LabelReview} \label{5-maj-review-analysis}
\begin{figure}[t]
	\centering    
	\subfloat[Data from development process.\label{fig:5maj_review}]{\includegraphics[width=0.23\textwidth]{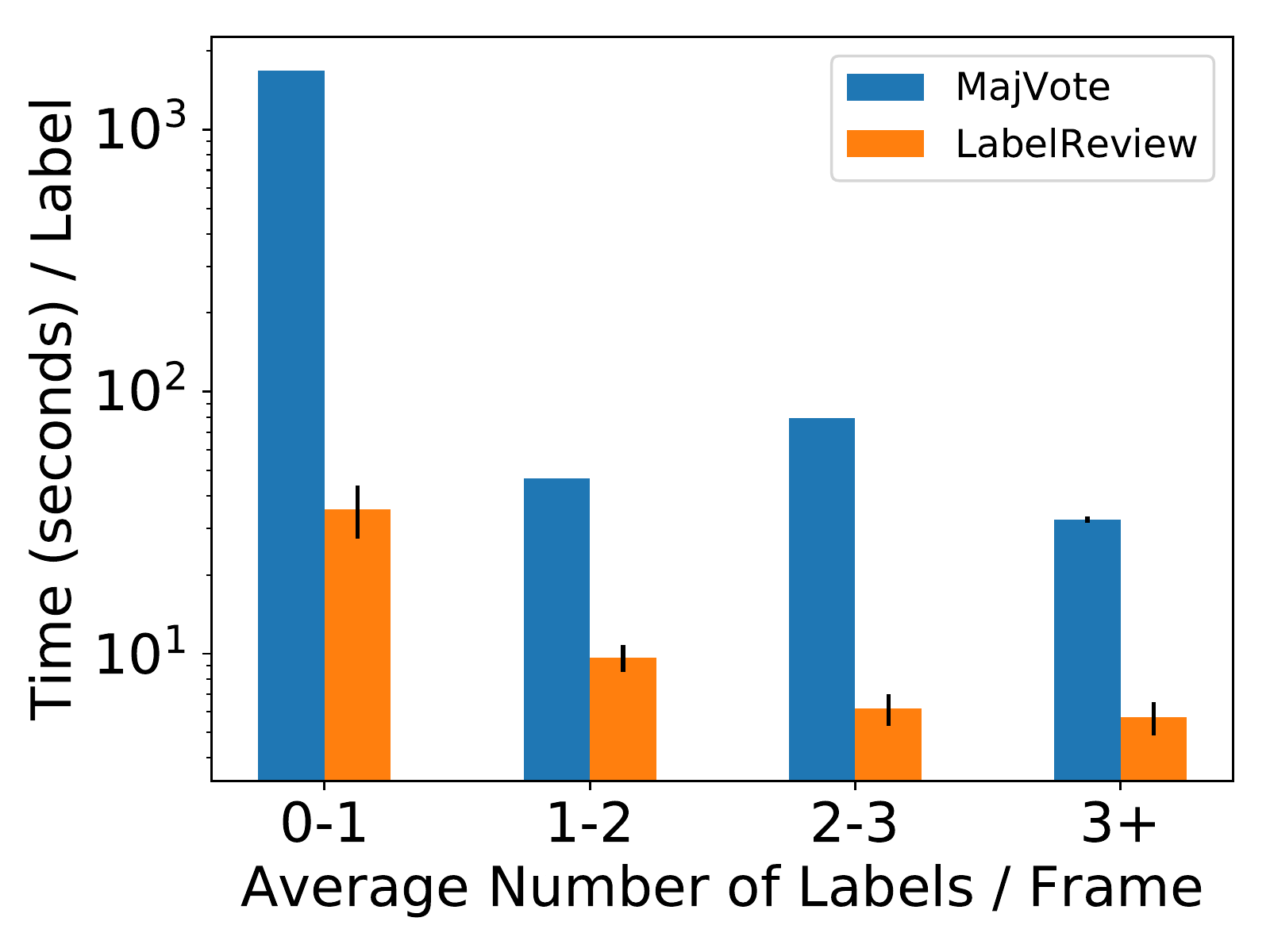}}\hfill
	\subfloat[Data from additional tests. \label{fig:exp_5maj_review}]{\includegraphics[width=0.23\textwidth]{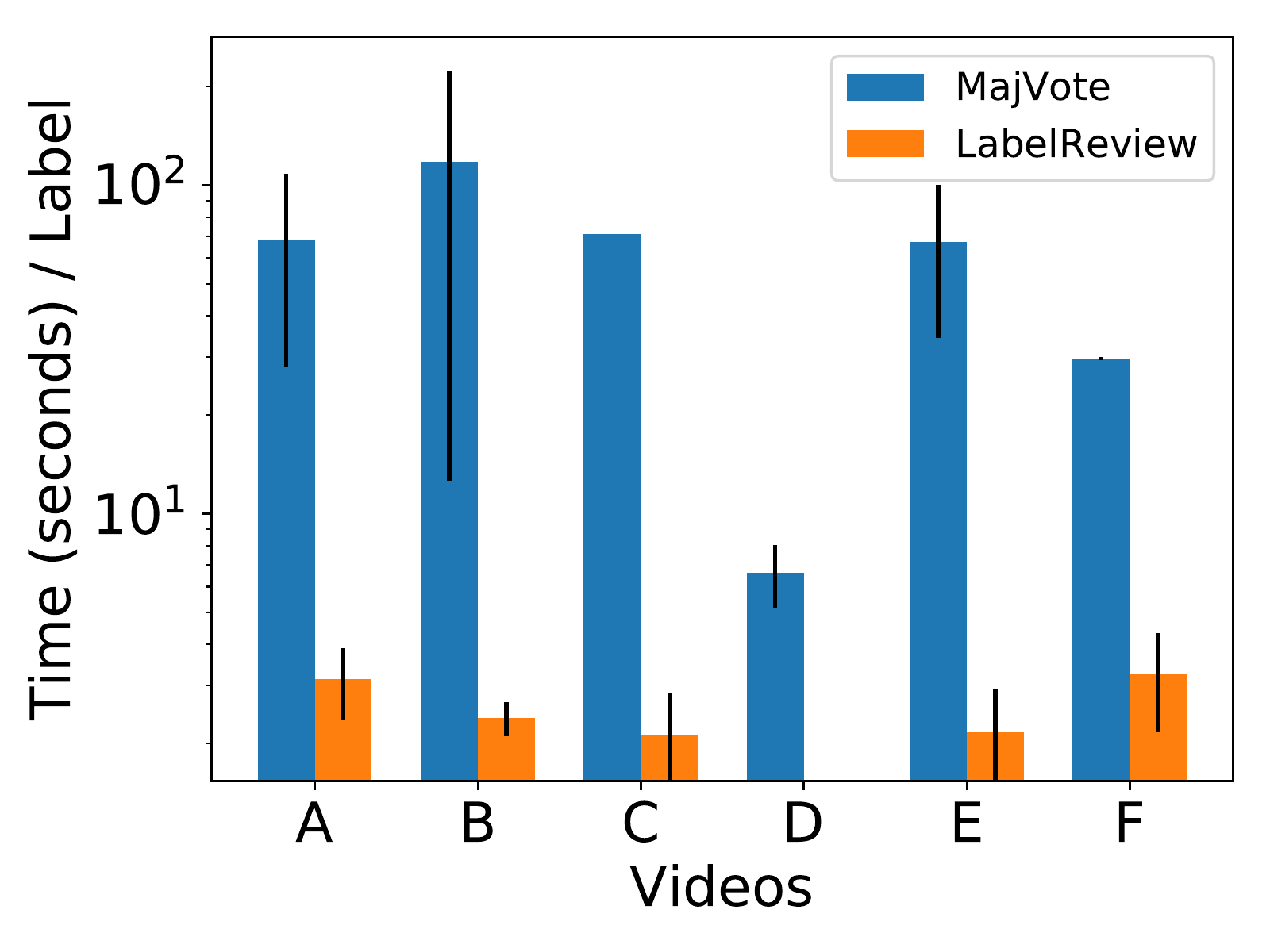}}\hfill
	\caption{Overall efficiency with different labeling frameworks.}
\end{figure}

Fig. \ref{fig:5maj_review} and Fig. \ref{fig:exp_5maj_review} show the comparison on overall efficiency between \MajVote~and \LabelReview. The total person time per final label is lower on average when we use \LabelReview, based on data collected through both the development process and additional tests.
From development, there are only seven videos for which we get final labels using \MajVote, two of which do not produce any consensus labels. There are more than 70 videos for which we get final labels through \LabelReview.

During additional tests, we tested two versions using \MajVote~and three versions using \LabelReview, which means the value of each bar is averaged over two or three samples. We exclude one sample for Video C where no consensus labels were achieved through \MajVote. The \LabelReview~efficiency for Video D is 0.63 with a standard error of 0.09.

\begin{figure}[t]
	\centering    
	\subfloat[For the seven videos with five sets of labels during development process.\label{fig:5maj_consensus}]{\includegraphics[width=0.23\textwidth, height=4cm]{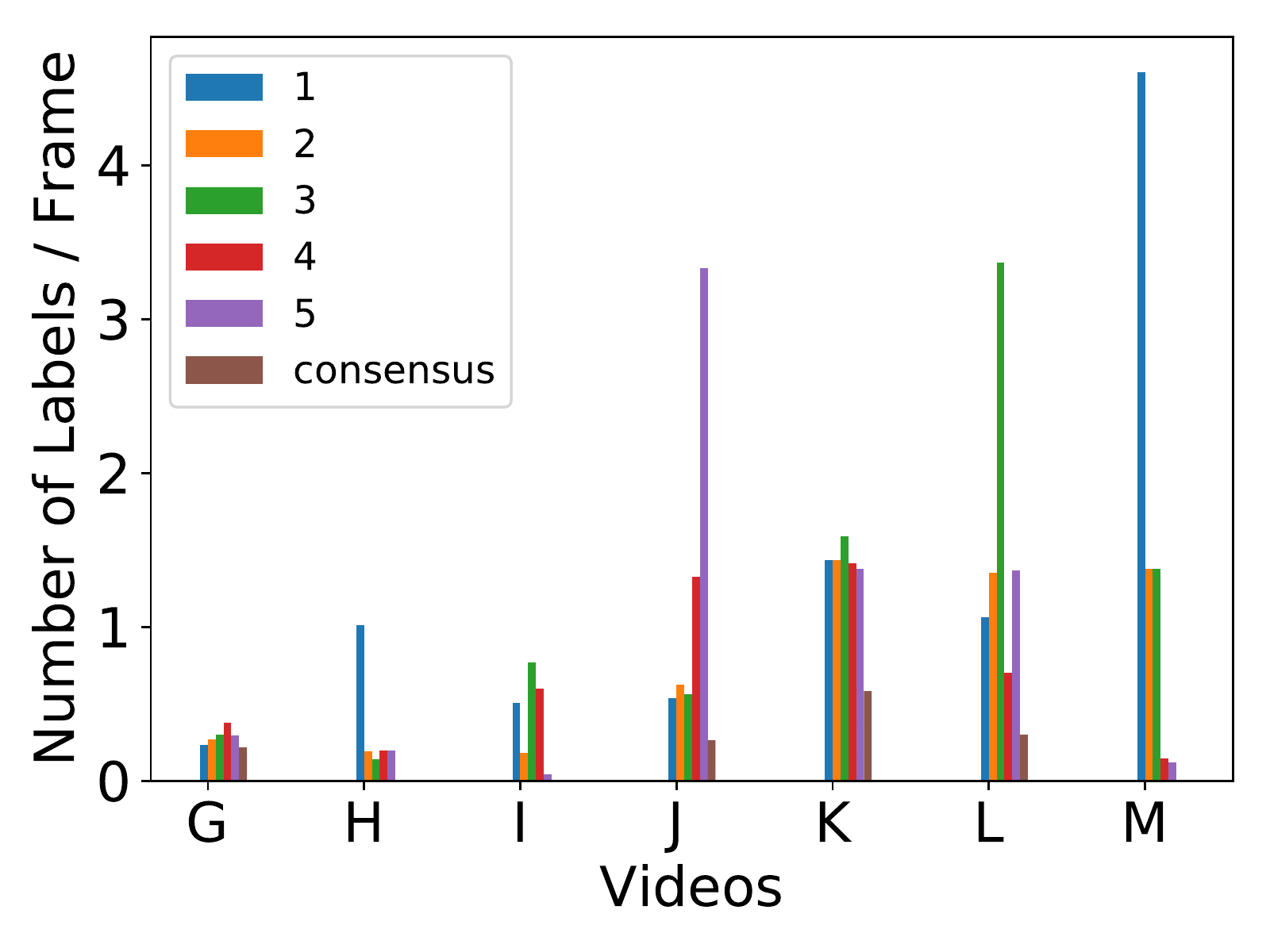}}\hfill
	\subfloat[For the six videos used in the additional tests under version \vBasic. \label{fig:exp_consensus}]{\includegraphics[width=0.23\textwidth, height=4cm]{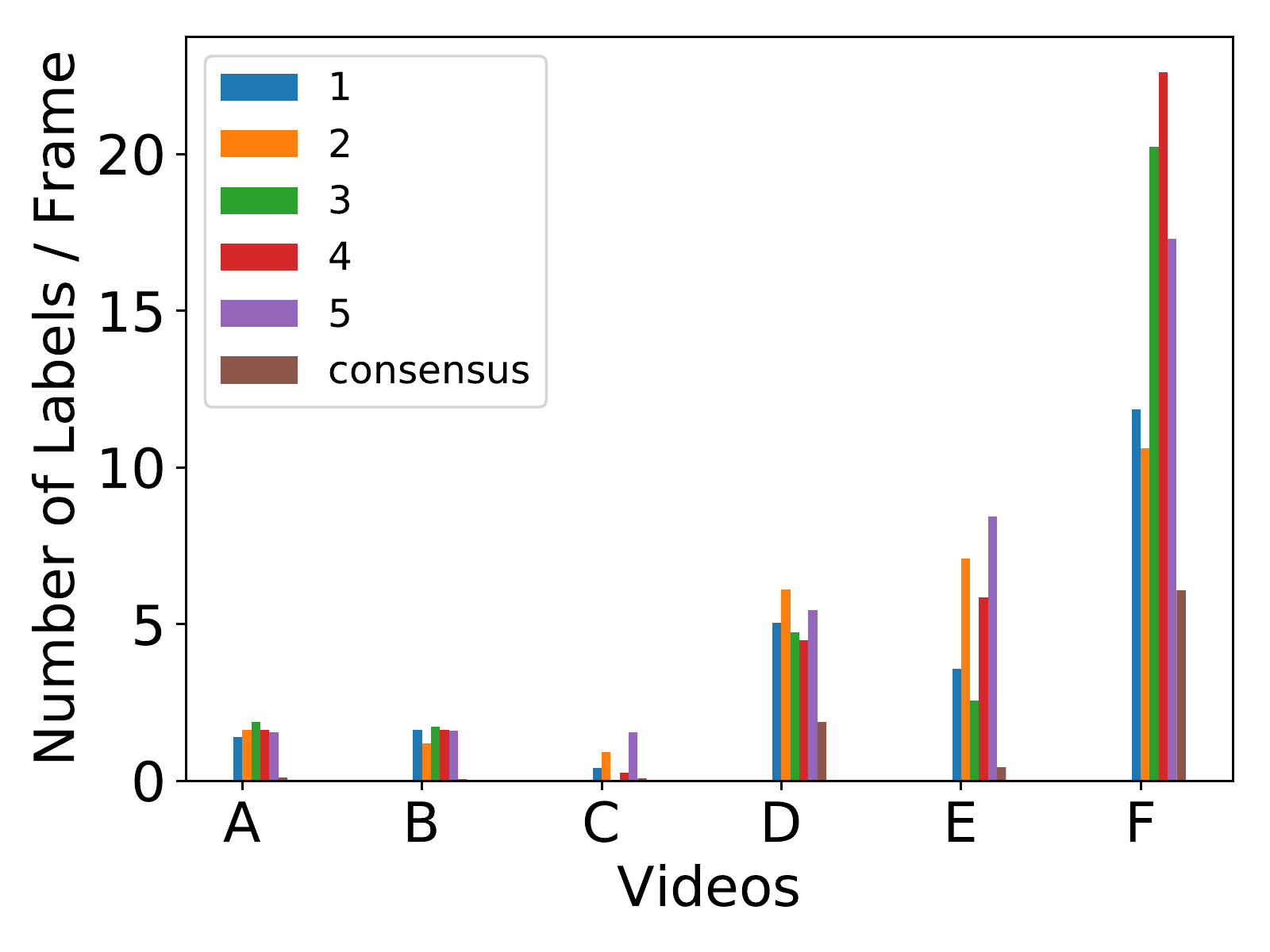}}\hfill
	\caption{Number of labels per frame for individual labelers and for consensus.}
	\label{fig:consensus}
	\vspace{-12pt}
\end{figure}

\begin{figure}[t]
	\includegraphics[width=0.30\textwidth]{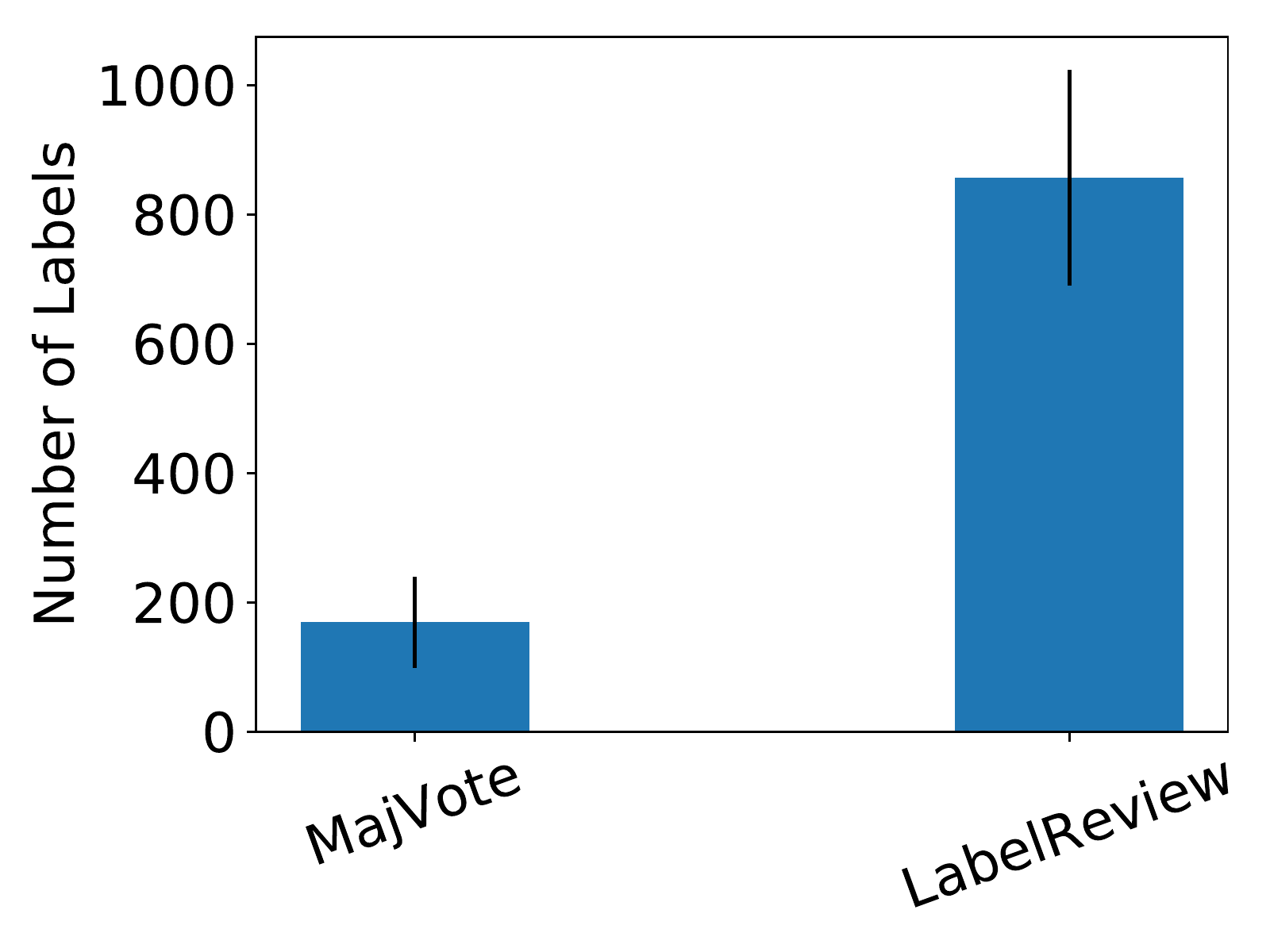}
	\centering
	\caption{Number of final labels for \MajVote~and \LabelReview~in additional tests.}
	\label{fig:experiment_num_labels_5maj_review}
	\vspace{-10pt}
\end{figure}

In addition to having more labelers involved, one reason that \MajVote~leads to a higher person time per final label is the lack of consensus. Fig. \ref{fig:consensus} shows that there were large discrepancies in the number of labels between individual labelers, which led to fewer consensus labels (zero in Videos I and M). Fig. \ref{fig:experiment_num_labels_5maj_review} shows that \MajVote~leads to fewer final labels than \LabelReview~in the additional tests as well. This indicates that using \LabelReview~framework can get us closer to the goal of exhaustive labels when compared to \MajVote.

\subsection{Impact of Other Changes}
In this section, we examine the individual and overall efficiency of each version of \toolname~to analyze the impact of the other changes made during the development of \toolname. For individual efficiency, we calculate person time spent per label for each individual labeler or reviewer, regardless of whether that label has been confirmed to be a final label.

\begin{figure}[t]
	\includegraphics[width=0.23\textwidth, height=4cm]{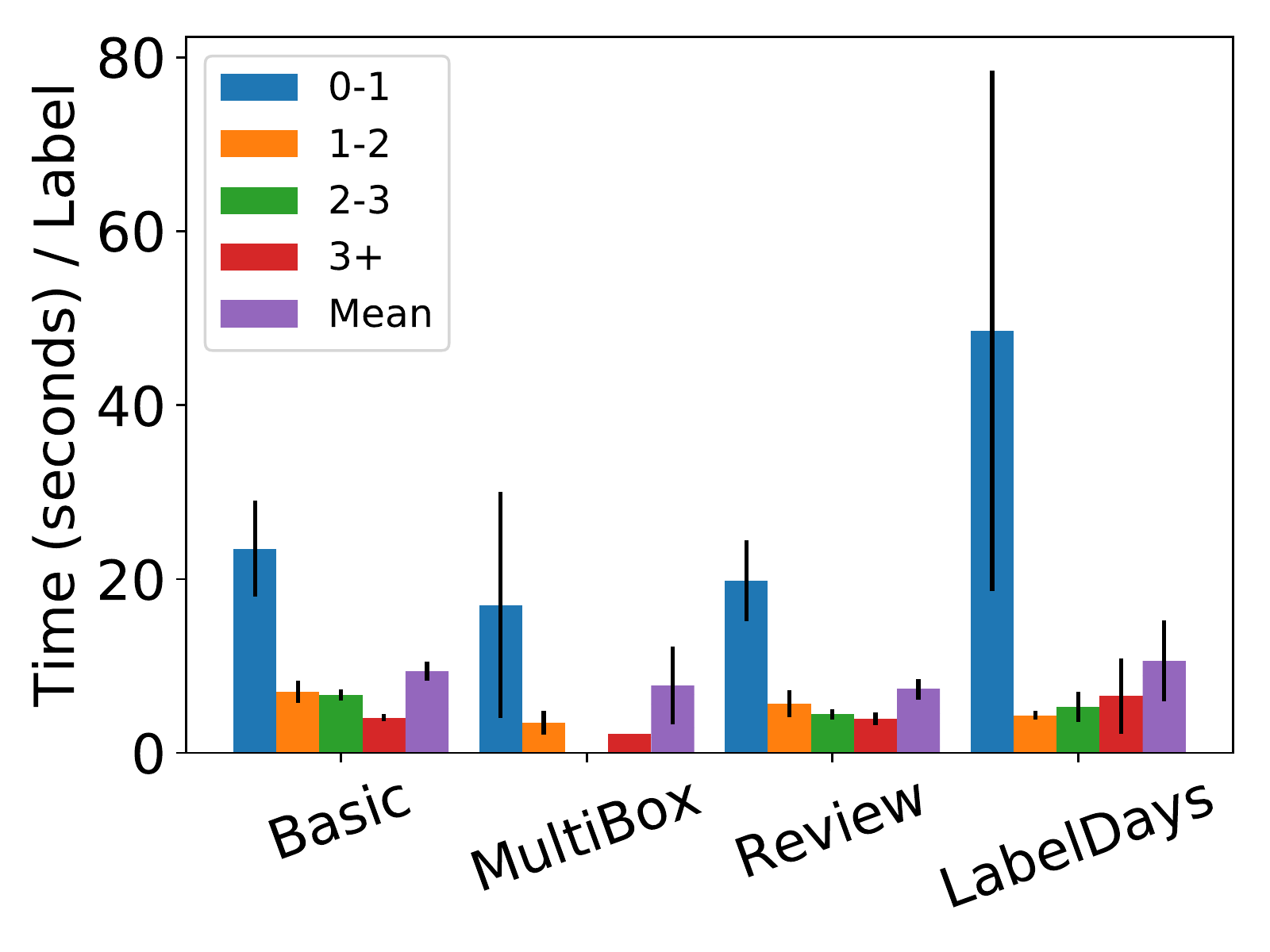}
	\includegraphics[width=0.23\textwidth, height=4cm]{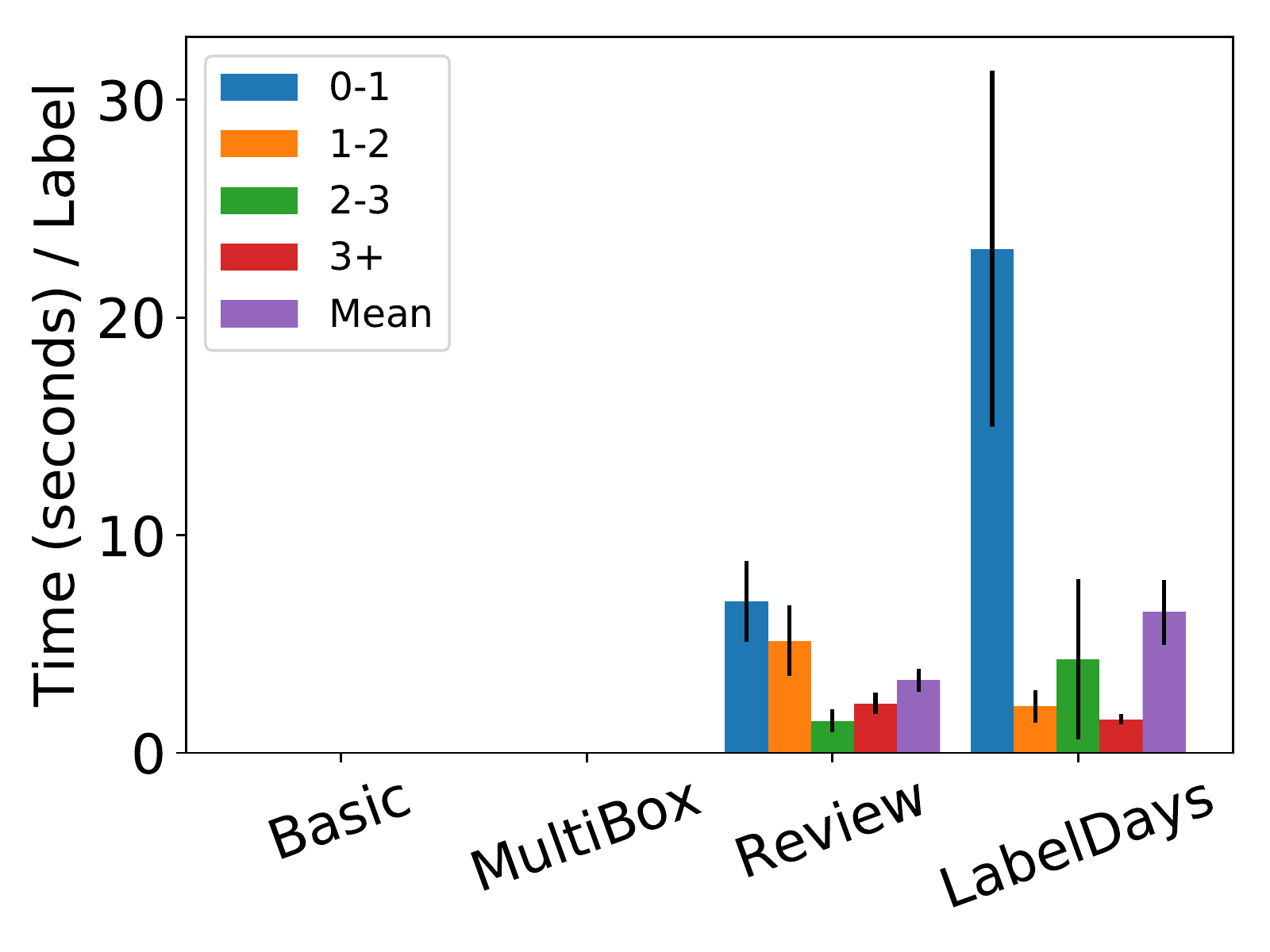}
	\centering
	\caption{Average individual efficiency of labeling (left) and review (right) with data collected during development.}
	\label{fig:mean_time_per_label}
	\vspace{-10pt}
\end{figure}

\begin{figure}[t]
	\includegraphics[width=0.23\textwidth, height=4cm]{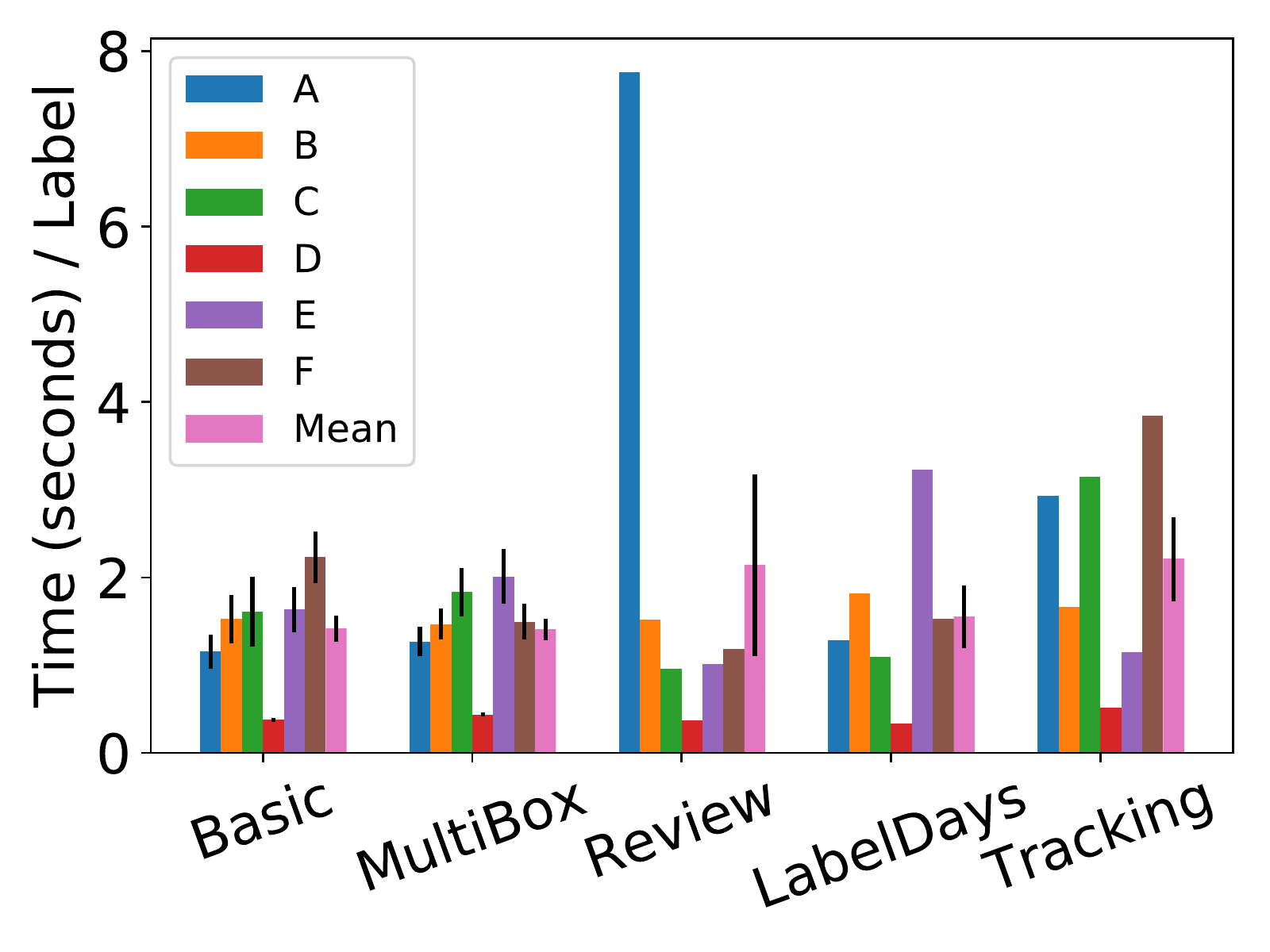}
	\includegraphics[width=0.23\textwidth, height=4cm]{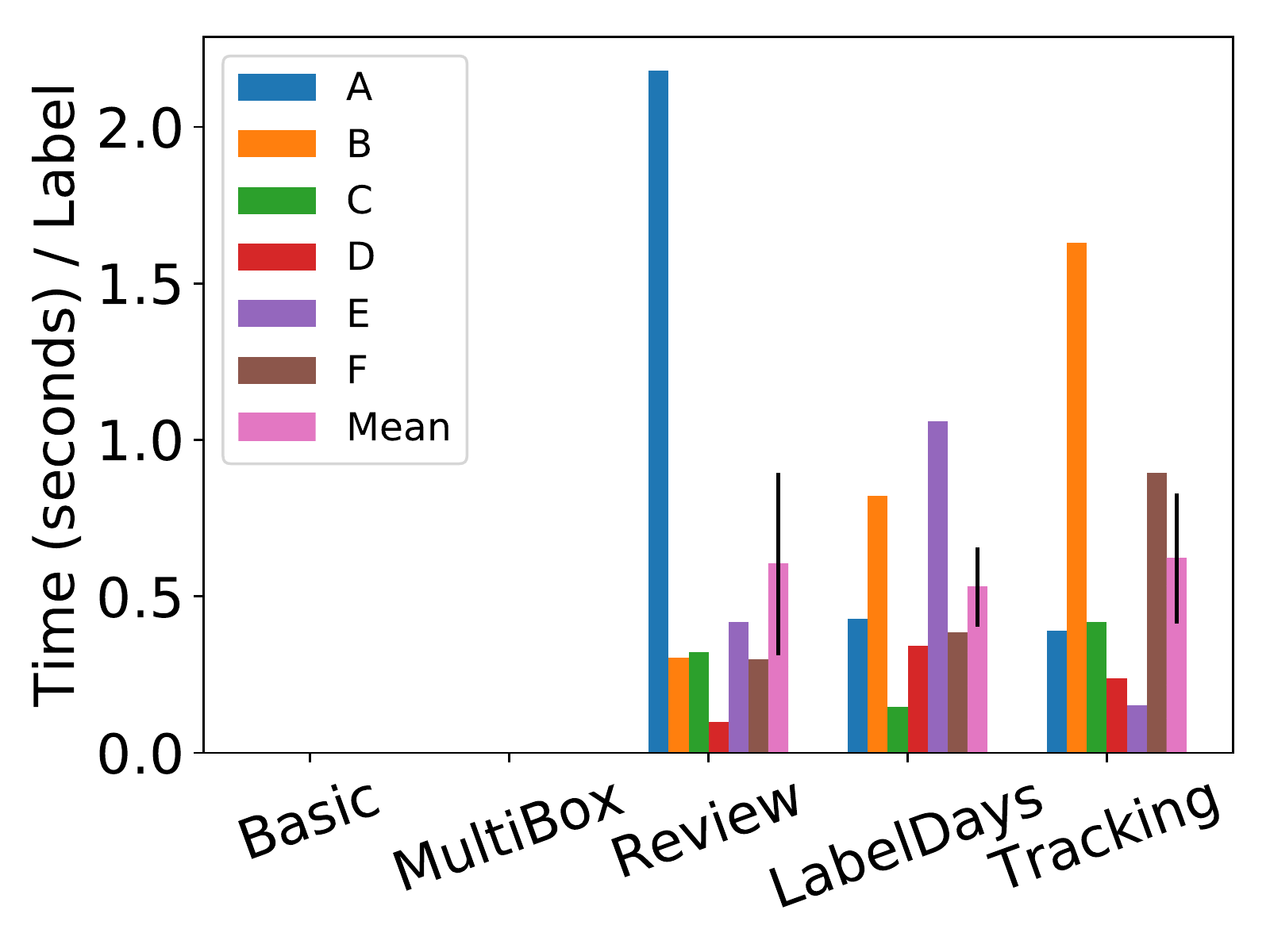}
	\centering
	\caption{Individual efficiency for each submission and average efficiency of labeling (left) and review (right) with data collected from additional tests.}
	\label{fig:scatter_experiment}
\end{figure}

We first show results of individual efficiency based on person time data collected during the development process in Fig. \ref{fig:mean_time_per_label}, which shows the mean labeling and reviewing time per label within the timespan of each change during development. We then examine the individual efficiency for labeling and reviewing in the additional tests (Fig. \ref{fig:scatter_experiment}). The results of each test have been shown by video, since there are only five sets of labels in the tests with \MajVote~(Version 1-2) and only one set of labels in the tests with \LabelReview~(Version 3-5). The five sets of labels in the \MajVote~tests are averaged by video, and the standard error bars are included. 
Fig. \ref{fig:scatter_experiment} shows that each change resulted in an improvement on the individual efficiency for some but not all of the videos.

\textbf{Multiple Box Selection}
The feature of multiple box selection was added to improve the individual efficiency of labeling. Checking the first two groups in Fig. \ref{fig:mean_time_per_label} and Fig. \ref{fig:scatter_experiment}, we notice that surprisingly, this feature improved individual efficiency for some of the videos (e.g., Video F), but not all the videos. One possible explanation is that in videos where there are many animals that did not move much over time, the changing position of the bounding boxes is mainly due to the movement of the camera. In this case, using multiple box selection is helpful. However, in other videos with only one or two animals in each frame, it may be faster to move the boxes separately, particularly if an animal moves.

\begin{figure}[t]
	\includegraphics[width=0.23\textwidth, height=4cm]{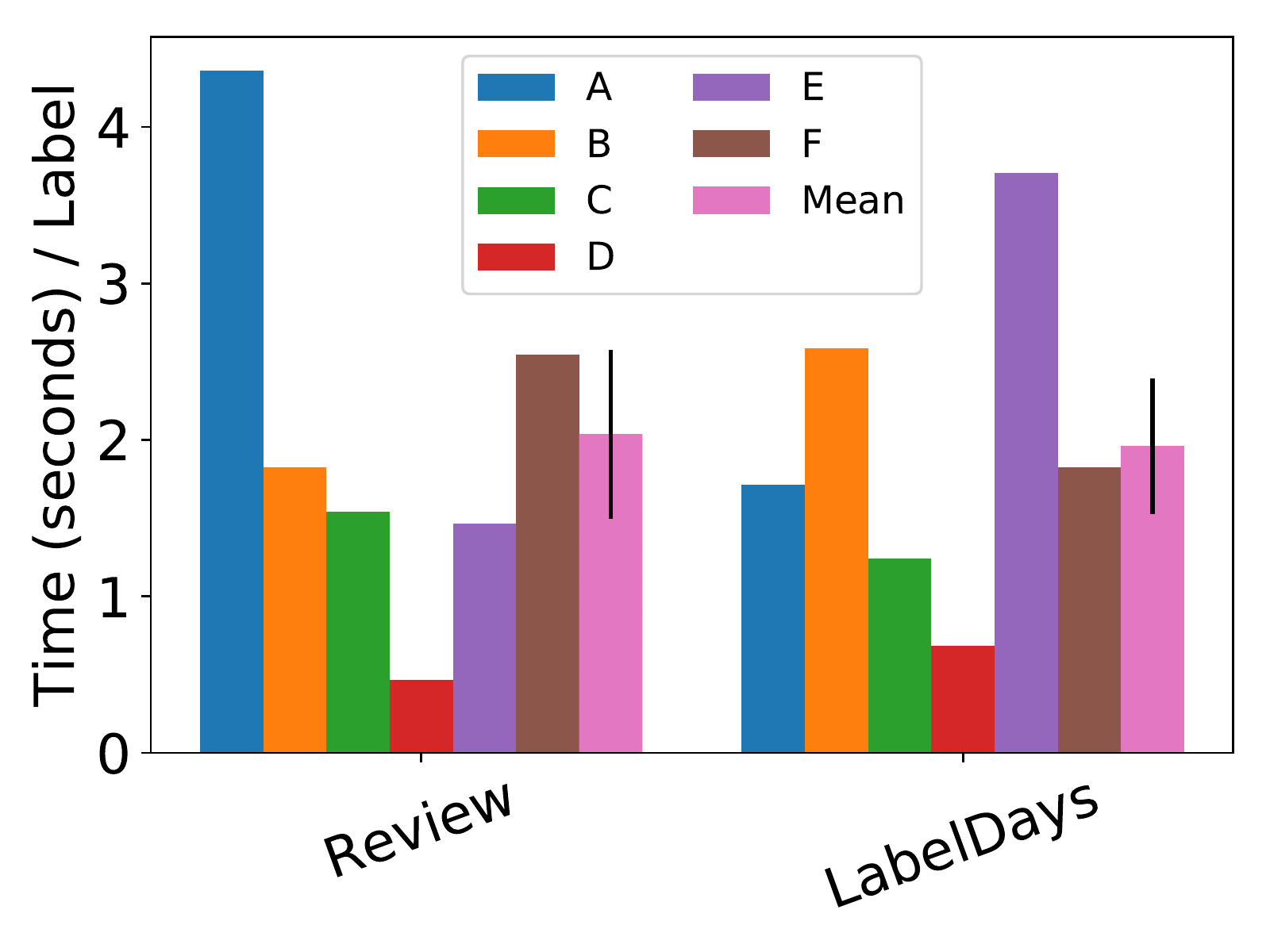}
	\includegraphics[width=0.23\textwidth, height=4cm]{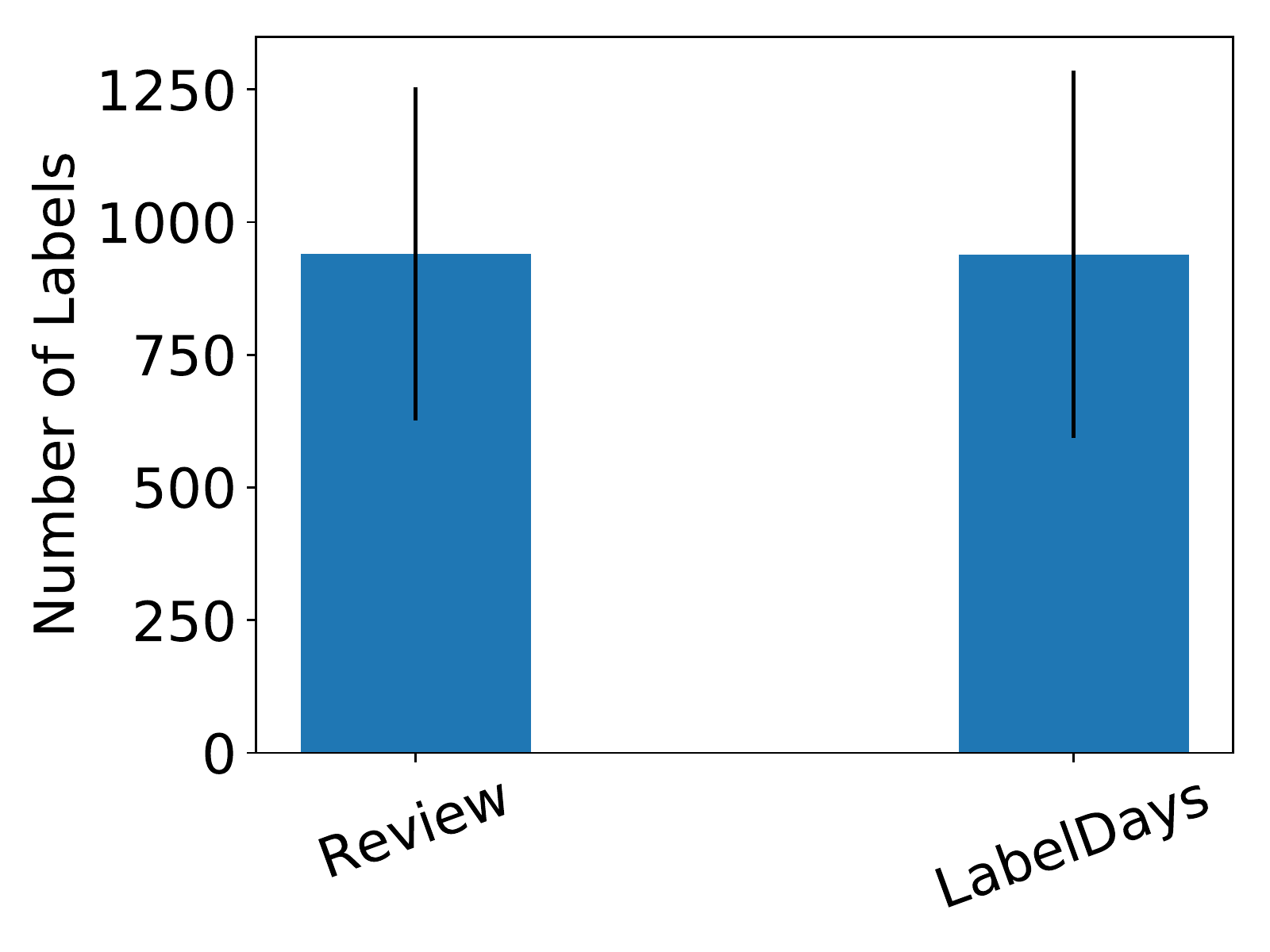}
	\centering
	\caption{Overall efficiency (left) and number of final labels (right) with \vReview~and \vLabelDays~during additional tests.}
	\label{fig:experiment_time_per_label}
	\vspace{-10pt}
\end{figure}

\textbf{Labeling Days}
Labeling days were introduced with the aim to increase the overall efficiency. Fig. \ref{fig:experiment_time_per_label} shows the average person time per final label has reduced from \vReview~to \vLabelDays~during additional tests, and the time per final label has reduced for Videos A, C, and F. Fig. \ref{fig:experiment_time_per_label} also shows the number of final labels has remained the same on average. 

The results indicate that introducing labeling days may help improving the efficiency and exhaustiveness of labeling, at least for some more complex videos. Subjective feedback from the labelers also indicated that introducing labeling days made it easier for them to deal with ambiguous cases, when it is difficult to maintain consistency and accuracy despite the guidelines.
 However, Fig. \ref{fig:mean_time_per_label} and Fig. \ref{fig:scatter_experiment} show that introducing labeling days does not lead to an improvement on individual efficiency in all cases. It is possible that it increased the individual labeling time due to extra discussion time, while saving time during review. We plan to analyze the effects of labeling days in more detail in the future. 

\textbf{Tracking}
We included the tracking feature in the additional test \vTracking~but it has not been deployed for the labelers to use. 
During the tests, we received positive feedback from labelers, particularly on videos in which animals were far apart and bright. 
In addition, the tracking feature was able to successfully track two animals in the first 10\% of Video B, as shown in Fig. \ref{fig:process}. Unexpectedly, the initial results from the additional test do not show a positive effect on time per label or number of labels. We believe this is due to the fact that it does not find a brightness threshold automatically, and is likely to track the wrong object when multiple objects are within the same search region. We plan to continue developing this feature given its promise in the cases where animals were far apart and bright. 

\textbf{Summary}
This section thus shows that while some of our proposed improvements led to increased efficiency, particularly the switch from \MajVote~to \LabelReview, in other cases (e.g., multiple box selection), surprisingly, it only increased efficiency in some videos. This result indicates that we must not add features on the intuition that they must improve performance, as they may only apply to some videos.

\section{Conclusions}

In conclusion, we presented \toolname, which provides a labeling and reviewing framework to gather labeled data from a small group of people in a secure manner, and a labeling interface with both general features for difficult video data, and specific features for our green security domain to track wildlife and poachers. We analyzed the impact of the framework and the features on labeling efficiency, and found that some changes did not improve efficiency in general, but worked only in particular types of videos. We plan to utilize the data we acquired in this work to estimate animal distributions, automatically detect wildlife and poachers in real-time, and predict poachers' movement patterns, which are important for game-theoretic approaches such as PAWS.
\section{Acknowledgments}

This research was supported by UCAR N00173-16-2-C903, with the primary sponsor being the Naval Research Laboratory (Z17-19598). It was also partially supported by the Harvard Center for Research on Computation and Society Fellowship and the Viterbi School of Engineering Ph.D. Merit Top-Off Fellowship.

\bibliographystyle{abbrv}
\bibliography{Gamesec2017}
\end{document}